\documentclass[aps,prb,amsmath,amssymb,superscriptaddress, reprint]{revtex4-1}

\usepackage{graphicx}

\begin{document}

\title{Polarization state of radiation from a photonic crystal slab}

\author{Chia Wei Hsu}
\affiliation{Department of Applied Physics, Yale University, New Haven, Connecticut 06520, USA}
\email{chiawei.hsu@yale.edu}

\author{Bo Zhen}
\affiliation{Research Laboratory of Electronics, Massachusetts Institute of Technology, Cambridge, Massachusetts 02139, USA}
\affiliation{Department of Physics and Astronomy, University of Pennsylvania, Philadelphia, Pennsylvania 19104, USA}

\author{Marin Solja\v{c}i\'{c}}
\affiliation{Research Laboratory of Electronics, Massachusetts Institute of Technology, Cambridge, Massachusetts 02139, USA}

\author{A. Douglas Stone}
\affiliation{Department of Applied Physics, Yale University, New Haven, Connecticut 06520, USA}

\begin{abstract}
We point out that the polarization state of radiation from a photonic crystal slab is strongly constrained by the direct non-resonant scattering process. 
The phase difference between the two linearly-polarized components in the far field can be predicted analytically and is largely independent of the periodic pattern.
We verify the prediction with full-field electromagnetic simulations.
\end{abstract}

\maketitle

Photonic crystal (PhC) slabs are structures periodic in the transverse directions and with a finite thickness in the third dimension~\cite{JJ_book, 1996_Krauss_Nature, 1999_Johnson_PRB, 2014_Zhou_PQE}.
The photonic modes are index guided at frequencies below the light line~\cite{JJ_book, 1996_Krauss_Nature,1999_Johnson_PRB},
while they couple to external radiations at frequencies above the light line~\cite{2014_Zhou_PQE, 2002_Fan_PRB, 2002_Tikhodeev_PRB}.
PhC slabs are attractive for their ease of fabrication and integration, tailorability, and their wide range of functionalities.
They have enabled high-power high-beam-quality lasers~\cite{2014_Hirose_NatPhoton} and many other devices~\cite{2014_Zhou_PQE},
and in recent years also became a versatile platform for studying emerging phenomena such as
bound states in the continuum~\cite{2013_Hsu_Nature, 2016_Gansch_LSA, 2017_Kodigala_Nature}
and non-Hermitian exceptional points~\cite{2015_Zhen_Nature}.


Polarization is one of the most fundamental properties of electromagnetic waves.
Controlling the polarization state is critical for optical communications, microscopy, nonlinear optics, and quantum optics, to name only a few.  
One can have the radiation from a PhC slab be in one of the two orthogonal linear polarizations by reducing the symmetry of the unit cell~\cite{2001_Noda_Science, 2002_Bristow_JQE, 2004_Lousse_OE, 2005_Altug_OL, 2007_Fedotov_PRL, 2008_Kilic_JOSAA, 2013_Huang_OE},
but it is limited to one of the two linear polarizations.
One can also control the beam pattern of radiation~\cite{2006_Miyai_Nature} or create vector beams~\cite{2006_Miyai_Nature, 2011_Iwahashi_OE, 2012_Kitamura_OL}, but at each outgoing angle the light is still mostly linearly polarized.
In fact, among the large body of literature on PhC slabs and related structures, only a few recent works indicate that the radiation can deviate from linear polarization~\cite{2011_Konishi_PRL, 2013_Shitrit_Science, 2014_Maksimov_PRB, 2015_Lobanov_PRB}.
To our knowledge, there has been no analytical treatment on the polarization state of resonant radiation from PhC slabs.
Here we show that part of the polarization state---specifically the phase difference between the two linear components---can be predicted with simple analytic theory and is largely independent of the actual periodic pattern.
Our results provide a useful tool for designing PhC slabs that emit with the desired polarizations.

\begin{figure}[b]
   \centering
   \includegraphics[width=\columnwidth]{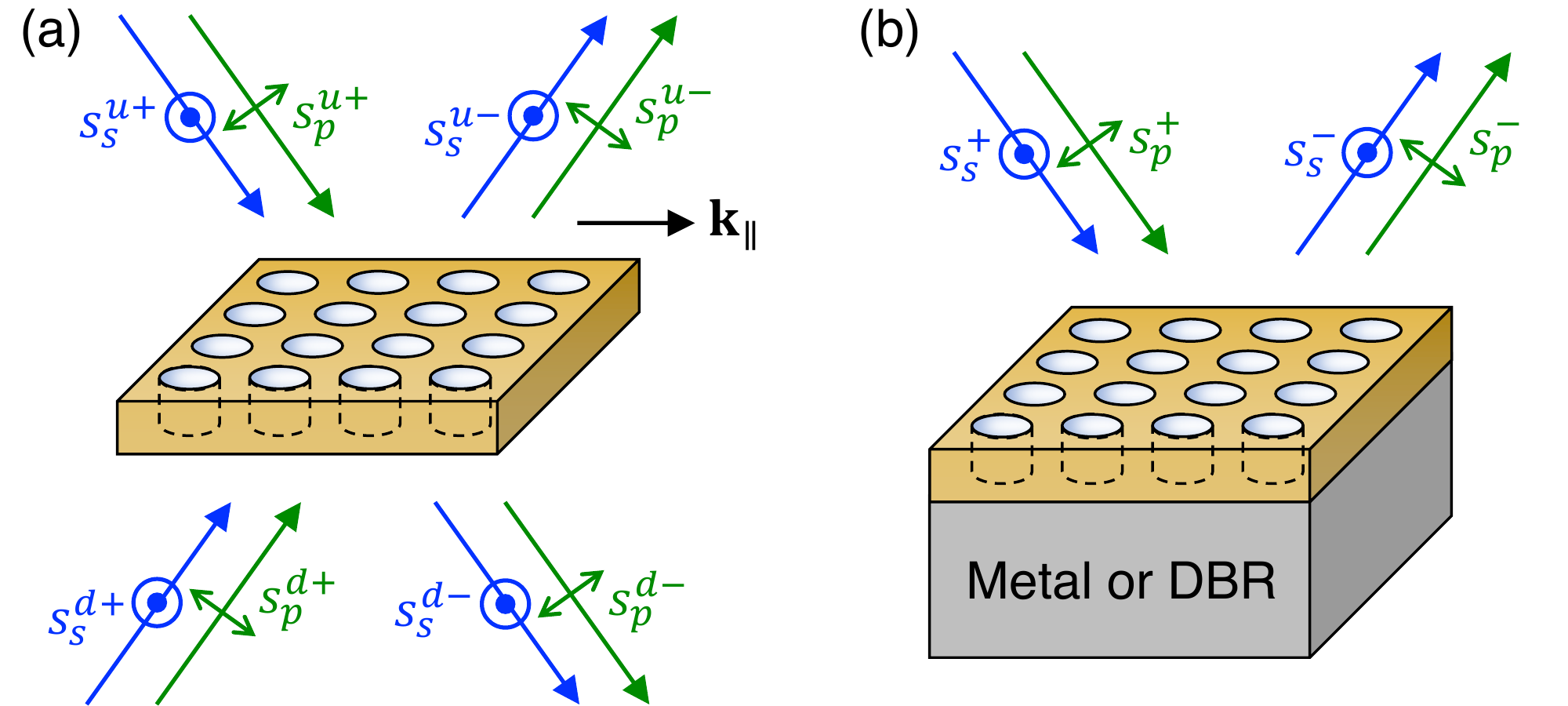} 
   \caption{Schematic illustration of 
   (a) two-sided and (b) one-sided PhC slabs
   and the corresponding input and output channels in both polarizations.
   }
\end{figure}

We start by considering a PhC slab that is periodic in $x$ and $y$ and has up-down symmetry ($\sigma_z$) and 180$^\circ$ rotational symmetry around the $z$-axis ($C_2$).
The dynamics of a resonance on an isolated band at in-plane wavevector ${\bf k}_{\parallel} = (k_x, k_y)$ can be described with temporal coupled-mode theory (TCMT)~\cite{Haus_book, 2003_Fan_JOSAA, 2004_Suh_JQE} via
\begin{align}
\label{eq:CMT_1}
{dA}/{dt}  &= \left( -i \omega_0 - \gamma \right) A + K^{\rm T} s^+, \\
\label{eq:CMT_2}
s^- &= C s^+ + D A, 
\end{align}
where $A$ is the resonance amplitude, $\omega_0$ its resonant frequency, and $\gamma$ the decay rate due to radiation.
Here we assume there is no gain or absorption loss.
The column vectors $s^{\pm}$ contain amplitudes of the incident ($+$) and outgoing ($-$) planewaves at ${\bf k}_{\parallel}$, in both $s$ and $p$ polarizations:
$s^{\pm} = (s^{u\pm}_s, s^{d\pm}_s, s^{u\pm}_p, s^{d\pm}_p)^{\rm T}$;
the superscripts $u$ and $d$ indicate above (up) and below (down) the slab respectively,
and we consider frequencies where there is no higher-order diffractions.
This is schematically illustrated in Fig.~1(a).
The amplitudes are normalized such that $|A|^2$ is the energy in the guided resonance, and $|s_{n}^{\pm}|^2$ is the longitudinal flux carried by the $n$-th planewave component.
Column vectors $K$ and $D$ contain coupling coefficients from the resonance to the four input/output channels:
$K = (k^{u}_s, k^{d}_s, k^{u}_p, k^{d}_p)^{\rm T}$,
$D = (d^{u}_s, d^{d}_s, d^{u}_p, d^{d}_p)^{\rm T}$.
The mirror symmetry $\sigma_z$ requires the field profile of the resonance to be either even or odd in $z$, so the coupling coefficients need to satisfy
\begin{equation}
\label{eq:sigma}
(d^{u}_s, d^{d}_s) = d_s (1, \sigma), \quad
(d^{u}_p, d^{d}_p) = d_p (1, \sigma), \quad
\sigma = \pm 1.
\end{equation}
When the transverse electric field $(E_x, E_y)$ is used to determine the phases of $A$ and $s^{\pm}$, we have $\sigma = 1$ and $-1$ for TE-like and TM-like resonances respectively.
Lastly, $C$ is a scattering matrix describing the direct (non-resonant) response and can be modeled as that of a homogeneous slab~\cite{2002_Fan_PRB, 2003_Fan_JOSAA}.
Since a homogeneous slab does not couple the two polarizations, $C$ is block diagonal with
\begin{equation}
\label{eq:Cs_Cp}
C = 
\begin{pmatrix}
C_s & 0\\
0 & C_p
\end{pmatrix},
\quad
C_s = 
\begin{pmatrix}
r_s & t_s\\
t_s & r_s
\end{pmatrix},
\quad
C_p = 
\begin{pmatrix}
r_p & t_p\\
t_p & r_p
\end{pmatrix}.
\end{equation}

When there is a finite amplitude $A$ in the resonance but no incident field ($s^+ = 0$), the outgoing radiation is $s^- = DA$; this can be the case when the resonance is excited via near field or fluorescence.
Since $A$ is an overall amplitude, the polarization state of the far-field radiation is uniquely defined by the coupling coefficients $D$.
For a high-$Q$ resonance, the mode profile does not change substantially when moderate gain is introduced, so the polarization state of the resulting laser emission is also given by $D$.
Such polarization state cannot be arbitrary because the the coupling coefficients are constrained by time-reversal symmetry.
Given $C_2$ symmetry, time reversal of the process $s^- = DA$ has incident field $(DA)^*$ with no outgoing field, so the directly scattered light $C(DA)^*$ must cancel the resonant radiation $DA^*$, meaning~\cite{2003_Fan_JOSAA, 2004_Suh_JQE}
\begin{equation}
\label{eq:CDD}
CD^* = -D.
\end{equation}
Eq.~\eqref{eq:CDD} may also be derived through energy conservation: unitarity of the resonant scattering process means that the resonant radiation must have certain phase relationship with the directly scattered waves.


Eq.~\eqref{eq:CDD} imposes a universal constraint on the possible polarization state of the resonant radiation.
Inserting Eqs.~\eqref{eq:sigma} and \eqref{eq:Cs_Cp} into Eq.~\eqref{eq:CDD} yields
\begin{equation}
\label{eq:d_phase}
{d_s^2}/{|d_s|^2} = - ( r_s + \sigma t_s)
\end{equation}
when $|d_s| \neq 0$, and similarly for the $p$ component.
Note that the right-hand side of Eq.~\eqref{eq:d_phase} has unit magnitude because of the unitarity of $C$.
Thus we have
\begin{equation}
\label{eq:arg_cs_over_cp}
\arg\left(\frac{d_s}{d_p}\right)
= \frac{1}{2} \arg \left( \frac{r_s + \sigma t_s}{r_p + \sigma t_p} \right) + N \pi,
\end{equation}
where $N$ is some integer.
That is, the phase difference between the two polarization components in the resonant radiation is determined uniquely (up to a possible $\pi$ phase difference) by the direct-scattering coefficients.
The direct-scattering coefficients do not depend on the actual pattern of the PhC slab, aside from some weak dependence on the effective index of the homogeneous slab and the resonant frequency.
Therefore, Eq.~\eqref{eq:arg_cs_over_cp} is universal.
Even though Eq.~\eqref{eq:CDD} has been known in the literature, its implication on the polarization state seems to be largely unnoticed until now.w
Given Eq.~\eqref{eq:arg_cs_over_cp}, the only free variable in the polarization state is $|d_s/d_p|$.

When the direct-scattering coefficients are the same for the two polarizations ($r_s=r_p$, $t_s=t_p$), $d_s/d_p$ needs to be real, so the resonant radiation is linearly polarized;
for isotropic materials, this is the case at the normal angle $(k_x=k_y=0)$ or at Fabry--P\'{e}rot resonances (where $t_s=t_p=\pm1$).
Away from these points, there will be a phase difference, and the resonant radiation is generally elliptically polarized unless $d_s=0$ or $d_p=0$ ({\it e.g.} along high-symmetry lines).
Approaching the grazing angle, the phase shift approaches $\pm \pi/2$ as the Fresnel reflection coefficients approach $r_p = -r_s = 1$.

Fig.~2(a) verifies the TCMT prediction of Eq.~\eqref{eq:arg_cs_over_cp} against finite-difference time-domain (FDTD) simulations~\cite{Taflove_book},
at different outgoing angles $\theta = \cos^{-1}(\hat{k}\cdot\hat{z})$ where $\hat{k}$ is the upward-pointing direction of radiation.
The simulated structure uses parameters from Ref.~\onlinecite{2013_Hsu_Nature} and is evaluated along a line $k_y = 0.5 k_x$ in the Brillouin zone; avoiding high-symmetry lines ensures that the radiating field generally contains both polarizations.
We calculate the two components of the far field, $d_s \propto \hat{s}\cdot{\bf E}_{\rm rad}$, $d_p \propto \hat{p}\cdot{\bf E}_{\rm rad}$, with $\hat{s} = \hat{z}\times\hat{k}_\parallel$, $\hat{p} = \hat{s}\times\hat{k}$, and ${\bf E}_{\rm rad}$ being the radiative field.
When the nodal line of $d_p$ is crossed as seem from the amplitudes in Fig.~2(b), we observe a $\pi$ phase jump in $\arg(d_s/d_p)$.
The only input for the TCMT prediction are the analytically known transmission and reflection coefficients of a homogeneous slab~\cite{2002_Fan_PRB, 2003_Fan_JOSAA},
yet the prediction matches the full-field simulations quantitatively.
Fig.~2(c) shows the degree of circular polarization,
$\rho_c \equiv (|d_{\rm R}|^2-|d_{\rm L}|^2)/(|d_{\rm R}|^2+|d_{\rm L}|^2)$,
where
$d_{\rm R} = (d_p + i d_s)/\sqrt{2}$ and
$d_{\rm L} = (d_p - i d_s)/\sqrt{2}$
are the far-field amplitudes in the circular polarization basis.

\begin{figure}[t]
   \centering
   \includegraphics[width=\columnwidth]{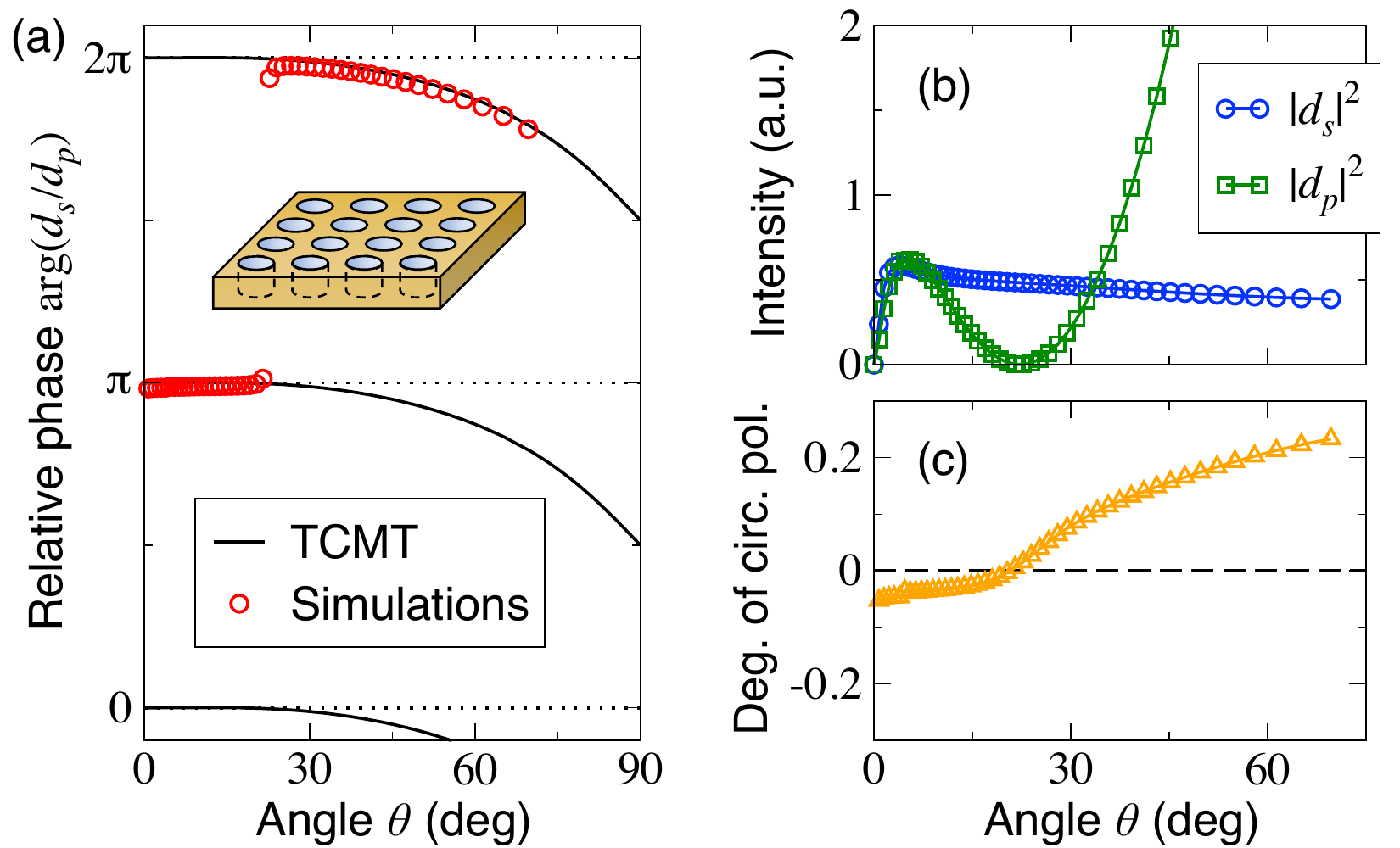}    
   \caption{
   (a) Relative phase between the two polarizations in the far field.
   Solid lines: TCMT prediction, Eq.~\eqref{eq:arg_cs_over_cp}.
   Symbols: FDTD simulations.
   Dotted lines indicate where the two are purely in- or out-of-phase.
   (b) Far-field intensities in the two linear polarizations.
   (c) Degree of circular polarization, $\rho_c \equiv (|d_{\rm R}|^2-|d_{\rm L}|^2)/(|d_{\rm R}|^2+|d_{\rm L}|^2)$.
   Simulation parameters are from Ref.~\onlinecite{2013_Hsu_Nature}, evaluated along $k_y = 0.5 k_x$ in the Brillouin zone.
   }
\end{figure}

Even though the far-field vector $\vec{d} = d_s \hat{s} + d_p \hat{p}$ may be elliptically polarized, it is still possible to define a linearly polarized vector $\vec{d'} = d_s' \hat{s} + d_p' \hat{p}$ via a smooth and deterministic phase projection,
\begin{equation}
d_s' \equiv d_s, \quad
d_p' \equiv d_p \sqrt{\frac{r_s + \sigma t_s}{r_p + \sigma t_p}}.
\end{equation}
Strictly speaking, the vector fields discussed in Ref.~\onlinecite{2014_Zhen_PRL} should have been defined via this linearly polarized vector $\vec{d'}$ rather than the possibly elliptically polarized vector $\vec{d}$.

With a slight modification, the above analysis can be generalized to PhC slabs without up-down mirror symmetry $\sigma_z$, but still in a uniform background ({\it i.e.}, same material above and below the slab).
The direct scattering matrix still takes on the form of Eq.~\eqref{eq:Cs_Cp}, but we no longer assume Eq.~\eqref{eq:sigma}.
By defining 
$d^{\pm}_s \equiv d^{u}_s \pm d^{d}_s$, 
$d^{\pm}_p \equiv d^{u}_p \pm d^{d}_p$, 
we get two equations
\begin{equation}
\label{eq:arg_cs_over_cp_C2_or_P}
\arg\left(\frac{d^{\pm}_s}{d^{\pm}_p}\right)
= \frac{1}{2} \arg \left( \frac{r_s \pm t_s}{r_p \pm t_p} \right) + N^\pm \pi
\end{equation}
with integers $N^\pm$.
With $\sigma_z$ symmetry, one of the two equations vanishes, and the other one reduces to Eq.~\eqref{eq:arg_cs_over_cp}.

This formalism also generalizes to PhC slabs without a 180$^\circ$ rotational symmetry $C_2$ (but still in a uniform background).
Since time reversal flips the propagation direction, both the resonance at ${\bf k}_\parallel$ and the resonance at $-{\bf k}_\parallel$ need to be considered~\cite{2013_Hsu_LSA, 2016_Zhou_Optica};
the two can have different properties in the absence of $C_2$.
This yields
$C_{{\bf k}_\parallel} D_{-{\bf k}_\parallel}^* = -D_{{\bf k}_\parallel}$, and
\begin{equation}
\arg\left[\frac{d_s^{\pm}({\bf k}_\parallel) d_s^{\pm}(-{\bf k}_\parallel)}{d_p^{\pm}({\bf k}_\parallel) d_p^{\pm}(-{\bf k}_\parallel)}\right]
= \arg \left( \frac{r_s \pm t_s}{r_p \pm t_p} \right),
\end{equation}
which jointly constrains the polarization state of radiation at ${\bf k}_\parallel$ and at $-{\bf k}_\parallel$.

The present formalism can also treat a PhC slab on a mirror substrate ({\it e.g.}, metal or distributed Bragg reflector) as shown schematically in Fig.~1(b);
we only need to set $d^d_{s,p}=0$ and $t_{s,p}=0$.
For example, we have
$\arg\left({d_s}/{d_p}\right) = [ \arg \left( {r_s}/{r_p} \right)/2] + N \pi$
when the structure is $C_2$ symmetric.

Note that the resonances of a homogeneous slab ({\it i.e.}, Fabry--P\'{e}rot resonances) must emit linearly polarized light because the two polarizations are decoupled; inhomogeneity is necessary for polarization mixing.
Meanwhile, for modes in a cylindrical fiber with non-zero propagation constant and non-zero angular momentum, the two polarizations are coupled at the circular dielectric boundary, so resonances in a fiber may emit elliptically polarized light even without inhomogeneity~\cite{2005_Schwefel_JOSAB}.

The present formalism provides a useful tool for predicting and engineering the polarization state of radiation from a photonic crystal slab.
It can be generalized to treat more complex geometries ({\it e.g.}, multiple layers, differing materials above and below, birefringent materials)
and to multiple resonances including the case of degeneracy.
One may also use it to study the polarization state of radiation from other types of resonators~\cite{2012_Ruan_PRA, 2014_Hsu_NL, 2012_Verslegers_PRL, 2014_Peng_nphys}. 

We thank Lei Shi, Ling Lu, and Yu Guo for helpful discussions.
C.W.H. and A.D.S. are partially supported by NSF via grants DMR-1307632 and ECCS-1068642.
B.Z. and M.S. are partially supported by the Army Research Office through the Institute for Soldier Nanotechnologies under contract no.~W911NF-13-D-0001 and by the Materials Research Science and Engineering Centers program of the NSF under award no.~DMR-1419807.


%

\end{document}